\documentclass[letter,twocolumn]{jpsj3}
\usepackage{txfonts}

\usepackage{epsfig}
\usepackage{graphicx}
\usepackage{dcolumn}
\usepackage{bm}
\usepackage{color} 
\usepackage{cite}
\usepackage{xspace}

\newcommand{\TN}{$T_{\rm N}$\xspace}

\newcommand{\Tz}{$T_0$\xspace}
\newcommand{\EF}{$E_{\rm F}$\xspace}
\newcommand{\cf}{$c$-$f$\xspace}

\newcommand{\CCS}{CeCoSi\xspace}
\newcommand{\LCS}{LaCoSi\xspace}

\newcommand{\OC}{$\sigma_1(\omega)$\xspace}
\newcommand{\R}{$R(\omega)$\xspace}

\newcommand{\hw}{$\hbar\omega$\xspace}

\title{
Observation of Electronic Structure Modification \\in the Hidden Order Phase of CeCoSi
}

\author{   
Shin-ichi Kimura$^{1,2,3}$\thanks{kimura.shin-ichi.fbs@osaka-u.ac.jp}, 
Hiroshi Watanabe$^{1,2}$, 
Shingo Tatsukawa$^2$, 
and Hiroshi Tanida$^4$
}

\inst{   
$^1$Graduate School of Frontier Biosciences, Osaka University, Suita, Osaka 565-0871, Japan \\
$^2$Department of Physics, Graduate School of Science, Osaka University, Toyonaka, Osaka 560-0043, Japan \\
$^3$Institute for Molecular Science, Okazaki, Aichi 444-8585, Japan \\
$^4$Liberal Arts and Sciences, Toyama Prefectural University, Imizu, Toyama 939-0398, Japan
} 

\recdate{
\today
}

\abst{  
\CCS with no local inversion symmetric crystal structure ($P4/nmm$) exhibits a phase transition of unknown origin (Hidden Order: HO) at about 12~K (\Tz) above the antiferromagnetic transition temperature (\TN$=9.4$~K).
The electronic structure change across \Tz was investigated with high-precision optical reflection spectroscopy. 
The optical spectrum changed from a typical metallic behavior above \Tz to a gap-like structure at around 15~meV below \Tz.
The gap-like structure was unchanged across \TN except for the narrowing of the Drude component of carriers due to the suppression of magnetic fluctuations.
This result suggests a slight change from the typical metallic electronic structure above \Tz to that with an energy gap near the Fermi level in the HO phase.
The change in electronic structure in the HO phase was concluded to be due to electron/valence instability.
}   

\begin{document}  

\maketitle


Materials without inversion symmetry have recently attracted much attention due to their topological phenomena~\cite{Moessner2021}.
The electric field generated by the non-inversion symmetry produces 
Weyl semimetallic electronic structures with strong spin-orbit interactions in the bulk~\cite{Bernevig2015,Armitage2018}
and giant Rashba effects on the surface of solids~\cite{Manchon2015}.
Such novel characteristic physical properties and electronic structures can be realized in materials with non-inversion symmetric crystal structures.

In rare-earth compounds with non-inversion symmetric / non-centrosymmetric crystal structures, fascinating physical properties have been observed mainly in superconducting properties:
A Cooper pair with a helical spin structure produces a superconducting state with a high upper critical field in global non-centrosymmetric materials CePt$_3$Si, CeCoGe$_3$, and CeRhIn$_3$.~\cite{Bauer2004,Measson2009,Kimura2007e}.
Also, a locally non-centrosymmetric material, CeRh$_2$As$_2$, shows a superconducting transition with changing the parity by applying magnetic field~\cite{Khim2021}.
Locally non-centrosymmetric Kondo semiconductors, Ce$M_2$Al$_{10}$ ($M =$ Ru, Os), show an antiferromagnetic transition with a tiny magnetic moment, namely hidden order (HO), due to an electronic instability~\cite{Nishioka2009,Kimura2011b,Kimura2011a}.

\CCS, the title compound, is one of the locally non-centrosymmetric rare-earth compounds.
It has a CeFeSi-type crystal structure ($P4/nmm$, $D^7_{4h}$, No.~129), 
in which there is no local inversion symmetry on the Ce site~\cite{Udc1970}.
Since the electronic specific-heat coefficient $\gamma$ is almost consistent with that of \LCS without $4f$ electrons, 
the $4f$ electrons are almost localized at low temperatures~\cite{Tanida2019}.
The N\'eel temperature (\TN) is 9.4~K 
and the first excited state of the $4f$ crystal-field splitting is about 100~K ($\sim 10$~meV) 
evaluated by the specific heat~\cite{Tanida2019} and inelastic neutron scattering (INS) measurements~\cite{Nikitin2020}.
Recently, HO, in which the order parameter has not been revealed yet, 
has been observed at $12-13$~K (\Tz) higher than \TN~\cite{Tanida2019,Hidaka2022}.
The origin is still unclear 
even though a theoretical work on the presence of odd-parity multipoles, originating from staggered antiferromagnetic and antiferro-quadrupole orderings, has been proposed~\cite{Yatsushiro2020}.
Experimentally, \Tz shifts to the high-temperature side, and the anomaly at \Tz becomes more pronounced by applying pressure, 
{\it i.e.}, 38~K at 1.5~GPa~\cite{Lengyel2013,Tanida2018}
and also by using magnetic field~\cite{Hidaka2022}.
With these results, some anomalies in the HO phase have been observed:
The symmetry of the Co site becomes lowering observed by NQR~\cite{Manago2021},
the crystal symmetry becomes lowering slightly detected by a synchrotron x-ray diffraction~\cite{Matsumura2022},
and the sign of the magnetic resistance changes at \Tz~\cite{Hidaka2022}.
These results suggest that the electronic structure changes at \Tz, 
but the direct evidence of the electronic structure modification in the HO phase has not been observed.

Optical measurements such as optical conductivity (\OC) and angle-resolved photoelectron spectroscopy (ARPES) are essential in studying HO.
In URu$_2$Si$_2$~\cite{Mydosh2011}, the most famous heavy-fermion material with HO, 
a pseudogap was reported in the infrared \OC spectra in the early stages of a series of studies~\cite{Bonn1988}.
In addition, in the HO phase of Ce$M_2$Al$_{10}$, a pseudogap was observed in the infrared \OC spectra only in a certain direction~\cite{Kimura2011b,Kimura2011a}.
These examples show that infrared spectroscopy is useful for observing slight changes in electronic structures.

In this Letter, we performed high-precision infrared spectroscopy to investigate whether a change in the electronic structure of \CCS occurs at \Tz. 
We observed the change in the reflectivity (\R) spectrum and the \OC spectrum at \Tz and also at \TN.
Below the photon energy (\hw) of about 300~meV, in addition to two optical phonon peaks at \hw$\sim20$ and 40~meV, 
a monotonic increase toward lower energies due to a Drude absorption of carriers and interband transitions of low-energy excitations appeared.
High-precision temperature dependence measurements revealed that 
the electronic structure near the Fermi level (\EF) slightly changes to a gap-like form below \Tz. 
At \TN, the relaxation time of the Drude parameter was discontinuously changed, implying the suppression of magnetic fluctuations below the antiferromagnetic transition.



Single-crystalline \CCS samples were synthesized by a Ce/Co eutectic flux method~\cite{Tanida2019}.
The \R spectra have been measured using the as-grown (001) plane, 
where the electric vector of light was set perpendicular to the $c$-axis ({\boldmath $E$}~$\perp c$).
Near-normal-incident \R spectra were acquired in a photon-energy range of 5~meV~--~1.5~eV at 6~--~300~K 
to obtain information on the temperature-dependent \OC spectra via the Kramers-Kronig analysis (KKA)~\cite{Kimura2013}.
To get the highly accurate \R spectra less than $\pm0.3$~\%, 
we adopted a feed-back positioning system combined with an {\it in-situ} gold evaporation method~\cite{Kimura2008}. 
For extrapolation of the low and high energy sides in KKA, we used the conventional methods of the Hagen-Rubens function for metallic \R spectra below the lowest accessible energy and measuring \R spectra up to 30~eV only at 300~K~\cite{fukui2014} connected with a free-electron approximation $R(\omega) \propto \omega^{-4}$ above the highest accessible energy~\cite{Dressel2002}.
Additionally, in order to clarify small changes in the \R spectrum at metallic reflectivity higher than 90~\% for the electronic structure modification at \Tz, we measured the relative \R change by temperature from 6~K (below \TN) to 20~K (above \Tz) with fixing the sample position.
With this method, we were able to capture changes in the \R spectrum with an accuracy better than 0.1~\%.



\begin{figure}
\includegraphics[width=0.45\textwidth]{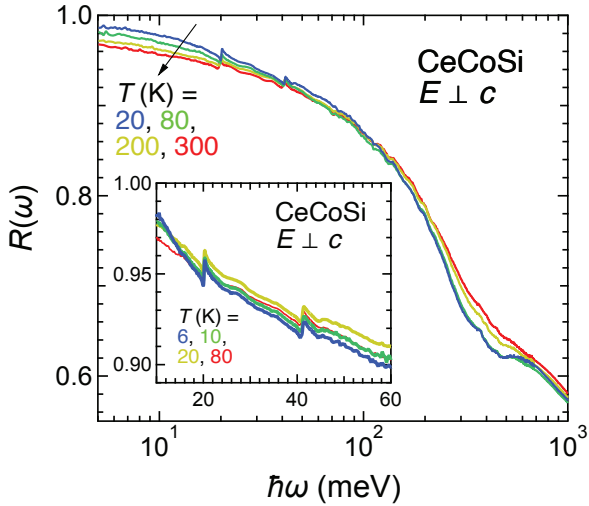}
\caption{
Temperature-dependent reflectivity (\R) spectra of the as-grown $(001)$ surfaces of \CCS in the photon energy \hw range of $5-1000$~meV.
Dispersive peaks at the photon energies \hw of about 20 and 42~meV originate from TO phonons.
(Inset) Temperature-dependent \R spectra of \CCS in the \hw range of $10-60$~meV at lower temperatures than 80~K.
The spectra at 6 and 10~K show a steep rise structure below 15~meV.
}
\label{fig:R}
\end{figure}

Obtained \R spectra of \CCS in a wide temperature range are shown in Fig.~\ref{fig:R}.
Although the overall structure is metallic, dispersive peaks due to optical phonons were observed at about 20 and 42~meV.
As the temperature is lowered, \R at $\sim10$~meV increases by about 2~\%, 
while it decreases by about 2~\% at $\sim300$~meV, 
indicating that the relaxation time increases at low temperatures, behaving like a typical metal.
In ordinary heavy-fermion systems, a double peak with a splitting energy of 0.25~eV corresponding to the Ce~$4f$ spin-orbit splitting width appears in this region, 
and a significant temperature dependence associated with the development of the Kondo effect is observed~\cite{Kimura2016a,Kimura2021b}.
However, in \CCS, there is no such change, suggesting that the $4f$ electrons are very localized, 
which is consistent with the speculation from the low electronic specific-heat coefficient~\cite{Tanida2019}.
The inset shows the temperature variation of the far-infrared \R spectrum below 80~K.
At 80 and 20~K, except for two phonon peaks, the backgrounds of the \R spectra monotonically increase toward lower energies.
However, the \R spectra at 10 and 6~K show a rapid increase with decreasing \hw below 15~meV.
This change suggests that a variation in the electronic structure appears below $\sim15$~meV.

\begin{figure}
\includegraphics[width=0.45\textwidth]{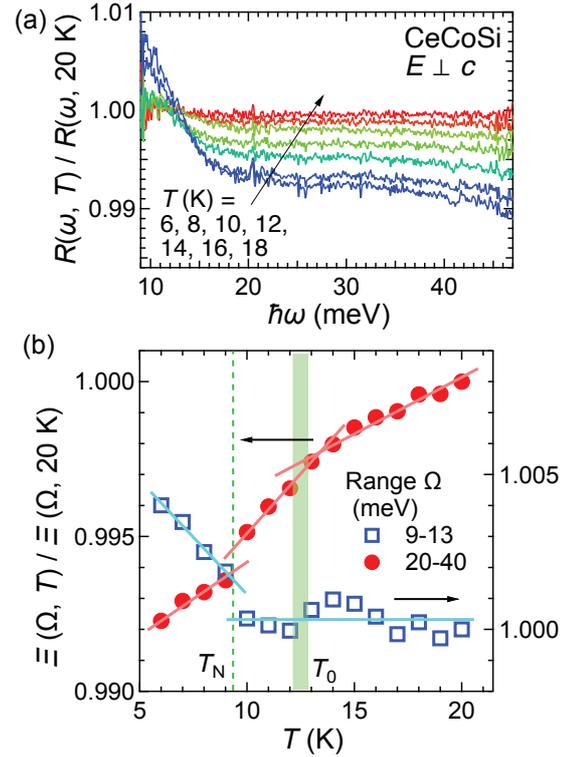}
\caption{
(a) Temperature-dependent \R spectra divided by that at 20~K ($R(\omega, T)/R(\omega, {\rm 20~K})$) of \CCS.
(b) Temperature dependence of the spectral integrations [$\Xi(\Omega, T)=\int_\Omega R(\omega, T)d\omega$] of the range $\Omega$ of 9--13~meV (open square) and 20--40~meV (solid circle) in (a), which are representative regions of the rapid increase below 15~meV and the gradual decrease above 15~meV, respectively, with decreasing temperature.
$\Xi(\Omega, T)$ is normalized by that value at $T=20$~K.
Solid lines are guides for the eye.
\Tz and \TN are shown by vertical thick-solid and dashed lines, respectively. 
}
\label{fig:intR}
\end{figure}

To examine the temperature variation of the \R spectrum below 20~K in detail, 
we measured the relative deviation to the \R spectrum at 20~K as shown in Fig.~\ref{fig:intR}(a).
The data show that the relative spectral change is about 0.1~\% and less, which is hardly detected using conventional \R measurements.
As observed in the inset of Fig.~\ref{fig:R}, 
the intensity above 15~meV decreases and conversely increases at lower energies with decreasing temperature.
To better clarify this change, we plotted the integration [$\Xi(\Omega, T)=\int_\Omega R(\omega, T)d\omega$]
normalized by the value at 20~K over the ranges $\Omega$ of 9--13~meV and 20--40~meV 
as a function of temperature in Fig.~\ref{fig:intR}(b).

$\Xi(\Omega, T)$ of $\Omega=9-13$~meV shows only a change of $\pm0.1$~\% or less above \TN, 
and is almost constant within the measurement accuracy of about $\pm0.1$~\%.
On the other hand, the value discontinuously changes at \TN and rapidly increases with decreasing temperature.
$\Xi(\Omega,T)$ of $\Omega=20-40$~meV also has a step-like change at \TN.
The spectral variation of $\Xi(\Omega, T)$ below \TN 
corresponds to an increase in the relaxation time of the Drude component 
and is consistent with the fact that the electrical resistivity drops below \TN~\cite{Tanida2019} due to a suppression of magnetic fluctuations by the magnetic ordering.

At around \Tz, $\Xi(\Omega,T)$ of $\Omega=20-40$~meV shows the increasing ratio with increasing temperature [$d\Xi(\Omega, T)/dT/\Xi(\Omega, {\rm 20~K})$] 
changes from $\sim 1 \times 10^{-3}$ K$^{-1}$ below \Tz to $\sim5 \times 10^{-4}$ K$^{-1}$ above \Tz. 
This result suggests that a modification of the electronic structure near \EF occurs at \Tz.
To clarify the electronic structure change in more detail, \OC spectra, which correspond to the joint density of states, were derived from KKA of \R spectra.

\begin{figure}
\includegraphics[width=0.45\textwidth]{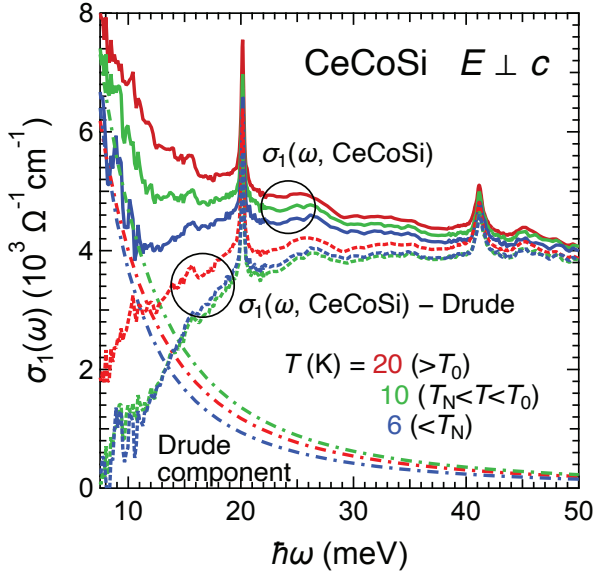}
\caption{
Optical conductivity (\OC) spectra of \CCS at representative temperatures of 20~K ($>$~\Tz), 10~K (\TN$<T<$~\Tz), and 6~K ($<$~\TN) (solid lines).
Dot-dashed lines are Drude curves obtained from the values of electrical resistivity and the \OC values at $\hbar\omega\sim5$~meV. 
The interband components evaluated by subtracting the Drude components from the \OC spectra are shown by dashed lines.
}
\label{fig:OC}
\end{figure}

Figure~\ref{fig:OC} shows \OC spectra at representative temperatures below \TN, between \TN and \Tz, and above \Tz to investigate the change of electronic states at the different phases.
Here, we obtained the \R spectra below 20~K by multiplying the relative \R spectra [$R(\omega,T)/R(\omega,{\rm 20~K})$] in Fig.~\ref{fig:intR}(a) to the \R spectrum at 20~K shown in Fig.~\ref{fig:R}, and derived \OC spectra from KKA of the \R spectra.
In this figure, the sharp peaks at 20 and 42~meV originate from optical phonons, 
and no evident change was observed in this temperature range.
This result suggests that there is no significant change in crystal structure or symmetry at \Tz and \TN, 
which is consistent with only a slight structural change observed with a synchrotron x-ray diffraction~\cite{Matsumura2022}.
Besides these sharp phonon peaks, the spectra have several broad peaks, possibly due to interband transitions.
However, they are not treated in this study because they hardly change in temperature.
At 20~K, the spectral intensity monotonically increases toward lower energies, 
which qualitatively represents the Drude component leading to direct current conductivity.
However, with decreasing temperature, 
it can be seen that the intensity around 15~meV decreases more rapidly than that above 20~meV.
This temperature dependence suggests a modulation in the electronic structure at low temperatures 
from a simple conduction band at 20~K to that with a gap-like structure.
The upward structure toward low energy below 15~meV becomes sharper at low temperatures, which reduces the intensity at around 15~meV.
Since the upward structure can be regarded as the Drude component originating from the carriers,
the Drude components expected from the electrical resistivity and the slope at the lowest accessible photon energy (dot-dashed lines) are subtracted from the \OC spectra as shown by dashed lines~\cite{Note}.
It should be noted that the Drude components in Fig.~\ref{fig:OC} were evaluated as no interband transition component at the lowest energy in \OC spectra.
This assumption may be unrealistic because interband transitions can exist even at low energy.
However, since the same evaluation method for the Drude components is used for all \OC spectra, the temperature dependence of the \OC spectrum after the Drude component subtraction can be compared with each other.

Compared to the Drude-subtracted \OC spectrum at 20~K ($>$~\Tz), 
the intensity below 20~meV is suppressed at lower temperatures than \Tz, indicating the appearance of an energy-gap-like structure.
The Drude-subtracted spectrum was almost unchanged across \TN from 10~K to 6~K.
This result suggests that the electronic structure near \EF is modified at \Tz, where an energy-gap-like form appears, but does not change at \TN.

\begin{figure}
\includegraphics[width=0.45\textwidth]{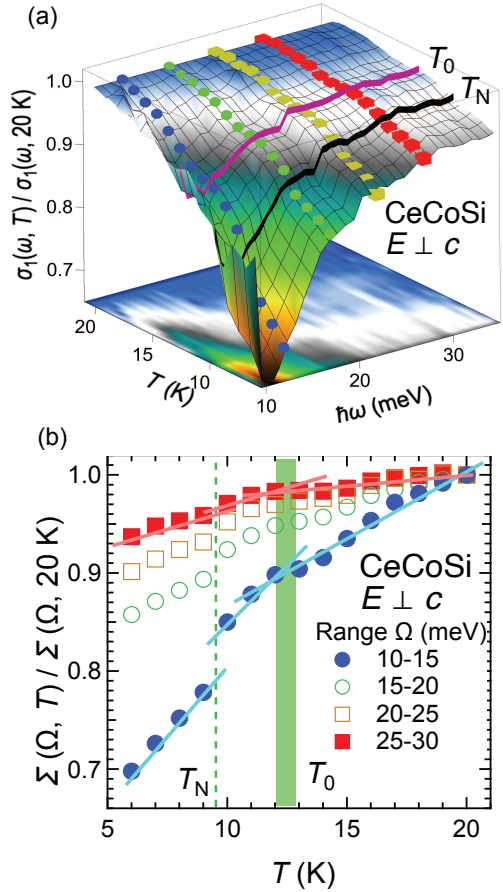}
\caption{
(a) Temperature-dependent \OC spectra normalized by that at 20~K of \CCS.
The marks are the same as those in (b).
(b) Temperature dependence of the spectral integrations [$\Sigma(\Omega, T)=\int_\Omega \sigma_1(\omega, T)d\omega$] of the range $\Omega$ of 10--15~meV (solid circle), 15--20~meV (open circle), 20--25~meV (open square), and 20--40~meV (solid square) in (a), 
which are representative regions of relative \OC spectra in (a).
The $\Sigma(\Omega, T)$ intensity at each $\Omega$ is normalized by the value at $T=20$~K.
Solid lines are guides for the eye.
\Tz and \TN are shown by vertical solid and dashed lines, respectively. 
}
\label{fig:relOC}
\end{figure}

To clarify the temperature dependence of \OC in more detail, 
the \OC spectra below 20~K are divided by that at 20~K and plotted as functions of temperature and photon energy in Fig.~\ref{fig:relOC}(a).
At low temperatures, the gap formation below 15~meV is clearly visible.
To clarify the spectral change at the ordering temperatures, the temperature dependence of the integration of the \OC spectrum 
[$\Sigma(\Omega,T)=\int_\Omega \sigma_1(\omega,T)d\omega$] 
in the four energy regions is shown in Fig.~\ref{fig:relOC}(b).
In $\Sigma(\Omega, T)$ of $\Omega=10-15$~meV, a clear jump appears at \TN, 
which is the same as the jump in \R at \TN in Fig.~\ref{fig:intR}(b).
This evidence can be attributed to the increase in the relaxation time of carriers caused by the suppression of magnetic fluctuations below \TN.
In addition, the increasing ratio of $\Sigma(\Omega, T)$ ($d\Sigma(\Omega, T)/dT$) below \Tz is steeper than that above \Tz.
This fact suggests the development of the gap-like structure below \Tz shown in Fig.~\ref{fig:OC}.
The intensity of $\Sigma(\Omega, T)$ at higher energies than 15~meV hardly decreases with decreasing temperature, confirming that the size of the energy gap is about 15~meV.
This gap-like structure does not change at \TN, which corresponds to the fact that $d\Sigma(\Omega, T)/dT$ is almost constant before and after \TN, even though a step-like change appears at \TN.
In other words, the gap-like structure observed at low temperatures in Fig.~\ref{fig:OC} starts to develop at \Tz.
This fact suggests that the HO of \CCS is accompanied by a modification in the electronic state near \EF.


Here, we discuss the origin of the electronic structure modification below \Tz.
An energy-gap-like structure with a gap size ($\Delta E$) of $\sim15$~meV slightly appears below \Tz.
The amount of the \OC intensity change at $\sim10$~meV below $\Delta E$ is less than 10~\% at 10~K (\TN$<T<$~\Tz)
compared to that at 14~K ($>$~\Tz),
and about 20~\% in maximum at 6~K ($<$~\TN) as shown in Fig.~\ref{fig:relOC}.
The change is much smaller than that of other HO materials; 
for instance, more than 50~\% in URu$_2$Si$_2$~\cite{Levallois2011} and CeOs$_2$Al$_{10}$~\cite{Kimura2011b}.
The reason is considered to be due to the remaining large background of the Drude component and low-energy interband transition that does not change with temperature.
This energy gap-like structure formation suggests that some bands close to \EF, including conduction bands, are modulated at \Tz.
The size of $\Delta E$ is similar to a broad peak observed by INS~\cite{Nikitin2020} and the $4f$ crystal-field splitting energy~\cite{Tanida2019}.
The optical gap observed here below \Tz may be related to both the INS peak and the crystal field, 
but we must consider that optical measurements detect the excitation at the momentum transfer of {\boldmath$q$}~$\sim 0$, in contrast to a finite {\boldmath$q$} detected by INS.
The origin of the optical gap and INS peak may be the \cf hybridization band because the evidence of a weak \cf hybridization appears in the electrical resistivity data~\cite{Tanida2019,Kawamura2022}.
Then, the valence of Ce-ions is slightly shifted from trivalent ($3+\delta$) above \TN.
On the other hand, below \TN, the Ce valence becomes trivalent because the $4f$ electrons in Ce are fully localized~\cite{Tanida2019}.
Therefore, at temperatures slightly above \TN, around \Tz, an electronic/valence instability can appear.
By applying pressure, the anomaly in the electrical resistivity at \Tz becomes more visible and moves to the high-temperature side, 
suggesting that the electronic/valence instability is grown with increasing pressure because the \cf hybridization intensity increases with applying pressure.
The pressure dependence typically appears in Ce-based heavy fermions~\cite{Yamaoka2015}.
The INS peak also shifts to the high-energy side with increasing pressure, which is the same as the behavior of the anomaly at \Tz, suggesting the enlargement of the \cf hybridization gap size.
Another possibility is that the electronic/valence instability may originate from the characteristic degeneracy specific to the nonsymmorphic crystal structure discussed in LaMnSi~\cite{Tanida2022b}.
Optical measurements under pressure may provide conclusive information for the origin of HO and its relation to the INS peak.

The observed electronic/valence instability at \Tz is similar to that of Ce$M_2$Al$_{10}$.
The crystal structure of Ce$M_2$Al$_{10}$ is also a locally non-centrosymmetric one ($Cmcm$), and the pressure-temperature phase diagram seems similar to that of \CCS~\cite{Umeo2011a}.
However, in Ce$M_2$Al$_{10}$, the lattice distortion at \Tz has yet to be observed, inconsistent with \CCS.
Like Ce$M_2$Al$_{10}$, in URu$_2$Si$_2$, the lattice distortion in the HO phase 
has not been observed~\cite{DeVisser1986,Niklowitz2010}.
Since all of these phase transitions are considered to originate from electronic/valence instability,
optical measurements, including ARPES, with high accuracy and resolution, will be essential in investigating the origin of HO.


To summarize, we investigated the origin of the HO phase of \CCS
using temperature-dependent high-precision relative infrared \R spectroscopy. 
The \R spectra and the \OC spectra after the Kramer-Kronig analysis clearly show non-monotonic temperature dependence at \Tz and \TN.
The anomaly at \TN originates from the elongation of the carrier relaxation time due to the suppression of magnetic fluctuations in the antiferromagnetic phase.
On the other hand, the change at \Tz is derived from the deviation of the low-energy interband transition component, with the emergence of an energy-gap-like structure with the size of $\sim15$~meV.
This result indicates that the electronic structure modification due to electronic/valence instability appears in the HO phase of \CCS.


\begin{acknowledgment}   
%
We would like to thank 
Mr. Kiichi Ikai for his support of the IR experiments, and UVSOR Synchrotron staff members for their support during synchrotron radiation experiments.
Part of this work was performed under the Use-of-UVSOR Synchrotron Facility Program (Proposals No.~22IMS6015) of the Institute for Molecular Science, National Institutes of Natural Sciences.
This work was partly supported by JSPS KAKENHI (Grant Nos.~20H04453 and 18H03683).

\end{acknowledgment}  

%
%

\bibliographystyle{jpsj}
\bibliography{../../../bibtex/library}

\end{document}